\documentclass[twocolumn,showpacs,preprintnumbers,superscriptaddress]{revtex4-1}

\usepackage{amsmath}
\usepackage{graphicx}
\usepackage{amsfonts}
\usepackage{amssymb}
\usepackage{bm}
\usepackage{color}
\usepackage{ulem}

\begin{document}

\title{\boldmath Dynamic Spin Fluctuations at $T\rightarrow 0$ in a Spin-$%
\frac{1}{2}$ Ferromagnetic Kagom\'{e} Lattice}
\author{Oren Ofer}
\affiliation{Schulich faculty of Chemistry, Technion - Israel Institute of Technology,
Haifa 32000, Israel}
\email{oren@physics.technion.ac.il}
\author{Lital Marcipar}
\affiliation{Department of Physics, Technion - Israel Institute of Technology, Haifa
32000, Israel}
\author{V. Ravi Chandra}
\affiliation{School of Physical Sciences, National Institute of Science
Education and Research, Institute of Physics Campus, Bhbaneswar, 751005
India}
\author{Snir Gazit}
\affiliation{Department of Physics, Technion - Israel Institute of Technology, Haifa
32000, Israel}
\author{Daniel Podolsky}
\affiliation{Department of Physics, Technion - Israel Institute of Technology, Haifa
32000, Israel}
\author{Daniel P. Arovas}
\affiliation{Department of Physics, University of California at San Diego, La Jolla, California 92093, USA}
\author{Amit Keren}
\affiliation{Department of Physics, Technion - Israel Institute of Technology, Haifa
32000, Israel}

\date{\today}
\begin{abstract}
We report magnetization, electron spin resonance (ESR), and muon spin
relaxation ($\mu $SR) measurements on single crystals of the $S=1/2$ (Cu$%
^{+2}$) kagom\'{e} compound Cu(1,3-benzendicarboxylate). The $\mu $SR is
carried to temperatures as low as 45~mK. The spin Hamiltonian parameters are
determined from the analysis of the magnetization and ESR data. We find that
this compound has anisotropic ferromagnetic interactions. Nevertheless, no
spin freezing is observed even at temperatures two orders of magnitude lower
than the coupling constants. In light of this finding, the relation between
persistent spin dynamics and spin liquid are reexamined.
\end{abstract}

\maketitle

The search for different kinds of quantum spin liquids (SLs) continues to
draw considerable experimental attention, and new candidate SLs are reported
from time to time \cite%
{Uemura:1994tz,FukayaPRL91,Bert:2004ij,Mendels:2007hz,Ofer:2006tj,Zorko:2008cc,Zorko: 2010jr,Fak:2012iu,Clark:2013ih}%
. Much of the search is focused on compounds with a kagom\'{e} lattice. SL
lack long range order and are classified according to the presence or
absence of a gap to magnetic excitations. The gapless ones, or those with
gap smaller than the lowest experimentally available temperature, are
expected to have persistent spin dynamics (PSD) at $T\rightarrow 0$. A major
experimental tool in the search for such states is the muon spin relaxation (%
$\mu $SR) technique. $\mu $SR is ideal for this task since it operates at
zero external field, without affecting the rotation symmetry of the
Hamiltonian. In addition, it can detect the presence or absence of long
range order, and dynamic fluctuations. Therefore, PSD has been frequently
used to identify materials as SL. However, $\mu $SR detected PSD in some
compounds that are not expected to be SL such as: pyrochlores \cite%
{DalmasdeReotier:2006kj}, molecular magnets \cite{SalmanPRB}, and other low
dimensional systems \cite{pratt}. This observation raises a question: can $%
\mu $SR give a false-positive observation when used to identify a SL?

\begin{figure}[tbh]
\begin{center}
\includegraphics[width=\columnwidth]{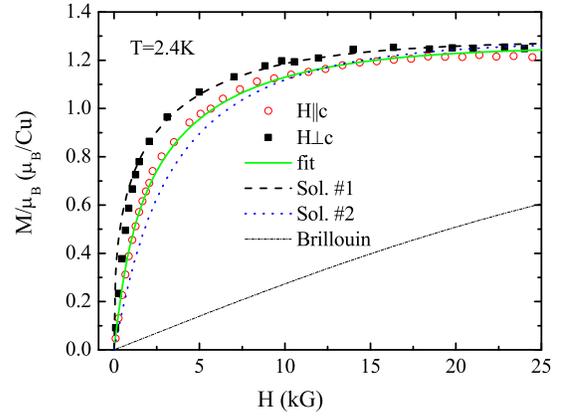}
\end{center}
\caption{(Color online) Magnetization of Cu(1,3-bdc) at $T=2.4$~K, measured
at two directions of the crystal: $\mathbf{H}\Vert \mathbf{\hat{c}}$ (black
filled symbols) and $\mathbf{H}\perp \mathbf{\hat{c}}$ (red hollow symbols).
The dashed-dotted line indicates a spin-$1/2$ Brillouin function with $%
g=2.0023$. The solid line indicates a fit to a Brillouin function with an
effective field (see Eq.~\protect\ref{Magnetization}). The dashed (dotted)
curves show the Brillouin function with an effective field using two
possible derived Hamiltonian parameters given in table~\protect\ref{tab:solu}%
.}
\label{fig:magnet}
\end{figure}

To address this question we investigate the organometallic hybrid kagom\'{e}
compound Cu(1,3-benzendicarboxylate) [Cu(1,3-bdc)], which was pointed out to
be a ferromagnet (FM) at low temperatures \cite{Nytko:2008ge}. Cu(1,3-bdc),
with the chemical formula CuC$_{8}$H$_{4}$O$_{4}$, has the ideal qualities
of a spin-1/2 kagom\'{e} featuring a non-magnetic 1,3-bdc ligand which links
the Cu$^{+2}$ kagom\'{e} layers \cite{Nytko:2008ge}. Initial magnetization
measurements on polycrystalline samples of Cu(1,3-bdc) suggested that the
mean nearest-neighbor super-exchange interaction is antiferromagnetic (AFM)
in nature with a Curie-Weiss (CW) temperature of $\Theta _{\text{CW}}=-33$%
~K, yet at low temperatures the onset of a FM signal was observed \cite%
{Nytko:2008ge}. Ferromagnetic correlation on a kagom\'{e} lattice means that
the degree of frustration is low, and therefore the spins should freeze at
low temperatures. Nevertheless, early $\mu $SR measurements showed only
slowing down of the spin fluctuations below $T_{s}=1.8$~K. The measurements
were carried out only down to $0.9$~K, where the magnetic state remains
dynamic with no long range order \cite{Marcipar:2009kh}.

Recently, single crystals were successfully synthesized in the form of
millimeter size flakes. Here we combine direction dependent bulk
magnetization and Electron Spin Resonance (ESR) measurements to characterize
the spin Hamiltonian of these crystals. We show that Cu(1,3-bdc) is an
anisotropic, slightly frustrated, ferromagnet; it is certainly not a SL. We
also extend the temperature dependence of the previous $\mu $SR measurements
and show that the dynamic fluctuations persist down to 45~mK. This result
indicates that PSD detected by $\mu $SR can give a false-positive when used
to identify a SL state.

The bulk magnetization ($M$) measurements were performed using a commercial
superconducting quantum interference device (SQUID) at temperatures $3\leq
T\leq 140$~K with external fields between $0.1\leq H\leq 25$~kG applied
along and perpendicular to the kagom\'{e} planes, i.e., $\mathbf{H}\parallel 
\mathbf{\hat{c}}$ and $\mathbf{H}\perp \mathbf{\hat{c}}$. The crystals were
held onto a small flat glass using epoxy glue with the $\mathbf{\hat{c}}$
direction perpendicular to the glass. The $\mathbf{\hat{c}}$ direction is
also perpendicular to the kagom\'{e} plane. The crystal's $\mathbf{\hat{a}}$
and $\mathbf{\hat{b}}$ directions are random. To determine the background
signal we measured the contribution from an identical glass with the epoxy
(not shown). This measurement indicated no temperature dependence and a
negligible background contribution compared to the sample signal.

\begin{figure}[tb]
\begin{center}
\includegraphics[width=\columnwidth]{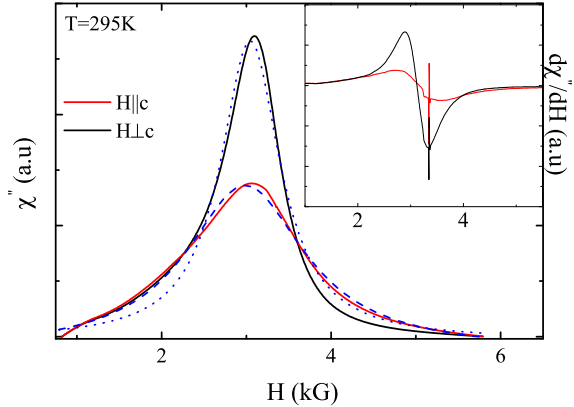}
\end{center}
\caption{(Color online) Representative ESR data for the two measured
directions, $\mathbf{H}\Vert \mathbf{\hat{c}}$ (red) and $\mathbf{H}\perp 
\mathbf{\hat{c}}$ (black), taken at $T=295$~K. The inset displays the ESR
raw signal. The main panel shows the integrated signal (absorption line).
The dashed lines demonstrates the fit to a Lorentzian function.}
\label{fig:esr}
\end{figure}

The magnetization measurements versus field at a temperature of $T=2.4$~K,
for two field directions, are plotted in Fig.~\ref{fig:magnet}. At fields
higher than about $15$~kG the magnetization saturates for both directions.
For $\mathbf{H}\perp \mathbf{\hat{c}}$, the saturation is reached at a lower
field than for $\mathbf{H}\parallel \mathbf{\hat{c}}$. This means that the
generated internal fields are strongest when the spins are in the kagom\'{e}
plane. The saturation value of the magnetization is $1.231(5)\mu _{B}$. This
suggests that the $g$ factor is higher than $2$. For a free spin $1/2$, the
field dependence of the magnetization $\mathbf{M=}g\mu _{B}\left\langle 
\mathbf{S}\right\rangle $ is given by the Brillouin function. This function
is plotted in Fig.~\ref{fig:magnet} by the dashed-dotted line. Clearly the
magnetization saturates at lower applied fields than expected for
non-interacting spins in both directions. This means that the internal field
is larger than the external one and that Cu(1,3-bdc) is a ferromagnet in our
experimental conditions.

To take interactions into account we consider a fully anisotropic
near-neighbours exchange Hamiltonian 
\begin{align}
\mathcal{H}=& \sum_{\left\langle i,j\right\rangle }\Big(J\mathbf{S}_{i}\cdot 
\mathbf{S}_{j}+DS_{i}^{z}S_{j}^{z}+E(S_{i}^{x}S_{j}^{x}-S_{i}^{y}S_{j}^{y}) 
\notag \\
& +F\left( \mathbf{S}_{i}\mathbf{\times {S}}_{j}\right) _{z}\Big)-g\mu
_{B}\sum_{i}\mathbf{S}_{i}\cdot \mathbf{H}.  \label{Hamiltonian}
\end{align}%
where the sum is over near-neighbours bonds, $J$ is the exchange coupling, $E
$ and $D$ are the anisotropies, and $F$ represents the $\hat{z}$ component
of the Dzyaloshinskii-Moriya (DM) term \cite{ElhahalPRB02}, which is often
the biggest \cite{ZorkoPRL08}. The $\mathbf{\hat{x}}$ direction is along
each bond, the $\mathbf{\hat{y}}$ direction is in the kagom\'{e} plane
perpendicular to each bond, the $\mathbf{\hat{z}}$ direction is
perpendicular to the kagom\'{e} plane. $D$, $E$, and $F$ are believed to be
due to spin-orbit couplings. $F$ is a first order and $E$ and $D$ are second
order effects. Nevertheless, $E$ and $D$ generate differences in the high
temperature magnetization between different directions, which, as we show
below, occur in our system. Therefore, $E$ and $D$ are certainly part of the
Hamiltonian \cite{Ofer:2009hy}. We start our analysis by assuming $F=0$,
and, as we shell see, there will be no reason to relax this assumption.

In the mean field approximation, the magnetization on each of the three kagom%
\'{e} sublattices $d$ is determined by the effective field this sublattice
experiences $\mathbf{H}_{\mathrm{eff}}^{d}$. This field is due to the
external field and the internal field generated by the moments $\mathbf{M}%
^{d}$ of the other sublattices, and is given by the generalized Brillouin
function 
\begin{equation}
\mathbf{M}^{d}=\frac{g_{\mathbf{\hat{H}}}\mu _{B}}{2}\tanh \left( \frac{g_{%
\mathbf{\hat{H}}}\mu _{B}}{2K_{B}T}\left\vert \mathbf{H}_{\mathrm{eff}%
}^{d}\right\vert \right) \mathbf{\hat{H}}_{\mathrm{eff}}^{d}
\label{Magnetization}
\end{equation}%
where $g_{\mathbf{\hat{H}}}$ represents the direction-dependent g-factor.
When the external field is in the $\mathbf{\hat{z}}$ direction, all
sublattices are magnetized in that direction only, their moments are equal,
and%
\begin{equation}
\mathbf{H}_{\mathrm{eff}}=[H-z(J+D)M_{z}\mathbf{/}(g\mu _{B})^{2}]\mathbf{%
\hat{z}}  \label{EffectiveHz}
\end{equation}%
regardless of $d$; $z$ is the number of neighbors. A solution of the
implicit Eqs.~\ref{Magnetization} and \ref{EffectiveHz} generates $%
M_{z}(H,J,D,g_{\Vert })$. We fit this $M_{z}$ to the $\mathbf{H}\parallel 
\mathbf{\hat{c}}$ data and find that 
\begin{equation}
J+D=-2.04(2)~K  \label{JplusD}
\end{equation}%
and $g_{\Vert }=2.51(1)$. The fit is plotted in Fig.~\ref{fig:magnet} by the
solid line. The calculated magnetization with interactions describes the
data quite well. This calculation demonstrates that the interactions in the $%
\mathbf{\hat{z}}$ direction must be ferromagnetic on order of $1$~K. 
\begin{figure}[t]
\begin{center}
\includegraphics[width=\columnwidth]{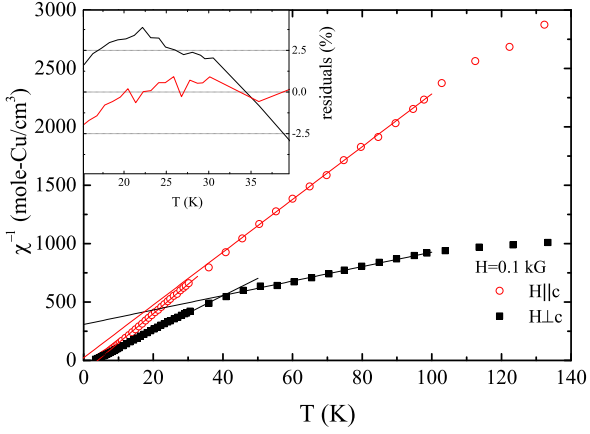}
\end{center}
\caption{(Color online) The temperature dependence of the inverse
susceptibility $\protect\chi ^{-1}(T)$, measured at two directions: $\mathbf{%
H}\Vert \mathbf{\hat{c}}$ (filled symbols) and $\mathbf{H}\perp \mathbf{\hat{%
c}}$ (hollow symbols), the solid lines are fits to a the inverse Curie-Weiss
law. The inset shows the residuals from the fit at low temperatures.}
\label{fig:suscep}
\end{figure}

Other Hamiltonian parameters are obtained from ESR. The ESR measurement were
done in the X-band ($\omega _{0}=9.5$~GHz) at $15\leq T\leq 300$~K. The
applied field was swept between $0.9\leq H\leq 6$~kG. The inset in Fig.~\ref%
{fig:esr} plots a representative raw ESR data taken at $T=295$~K of the
sample with a DPPH reference. To obtain the absorption line, we subtract the
reference signal, and integrate the raw ESR signal over the applied field.
The main panel of Fig.~\ref{fig:esr} shows the absorption lines for the two
measured directions. A reasonable fit to the absorption line is found to a
Lorenzian function, 
\begin{equation}
\chi ^{\prime \prime }(H)=\frac{A_{\Vert ,\bot }}{\pi }\frac{\delta }{\delta
^{2}+(H-H_{\Vert ,\bot })^{2}}  \label{eq:loren}
\end{equation}%
where $2\delta $ is the full width-half maximum and $H_{\Vert ,\bot }=\omega
_{0}/(g_{\Vert ,\bot }\mu _{B})$ is the resonance field. We find that $%
g_{\perp }=2.164(2)$ and $g_{||}=2.181(2)$, $\delta _{\perp }=0.432(1)$~kG
and $\delta _{||}=0.867(22)$~kG. The $\delta $ and $g$-factor do not have
temperature dependence down to $15$~K. The area $A$ of the $\mathbf{H}\perp 
\mathbf{\hat{c}}$ measurement is highest, consistent with the magnetization
data, and increases upon cooling as expected. The ESR $g_{\Vert ,\bot }$
factors are larger than $2$ but lower than the value determined by the
magnetization measurement. The cause of the discrepancy between the
magnetization and ESR $g$ factors is not clear to us.

\begin{figure}[tbp]
\begin{center}
\includegraphics[width=\columnwidth]{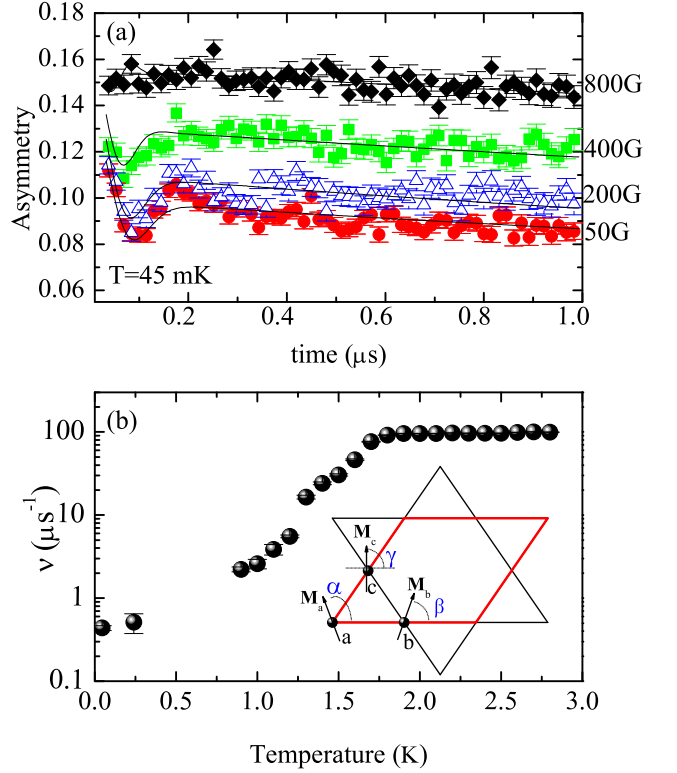}
\end{center}
\caption{(Color online) (a) The raw $\protect\mu $SR data with applied
longitudinal field of $50$ to $800$~G, the solid lines demonstrate the fit
to Eq.~\protect\ref{eq:kt}. (b) A semi-log scale of the spins fluctuation
rate $\protect\nu $ obtained in Ref.~\protect\cite{Marcipar:2009kh} with the
added points at 240~mK and 45~mK. Inset: the kagome unit cell (bold red)
with its three sublatices (a,b and c) the corresponding magnetic moment ($%
\mathbf{M}^{a}$, $\mathbf{M}^{b}$ and $\mathbf{M}^{c}$) and angles ($\protect%
\alpha $, $\protect\beta $, $\protect\gamma $).}
\label{fig:musr}
\end{figure}

At temperature higher than typical interaction strength, the widths $\delta
_{||}$ and $\delta _{\perp }$ are related to the three Hamiltonian
parameters via the moments according to 
\begin{equation}
g_{_{\Vert ,\bot }}\mu _{B}\delta _{\Vert ,\bot }=\frac{\pi }{\sqrt{3}}%
M_{2}^{\Vert ,\bot }\sqrt{\frac{M_{2}^{\Vert ,\bot }}{M_{4}^{\Vert ,\bot }}},
\label{delta}
\end{equation}%
where $M_{2}^{\Vert ,\bot }=-Tr\left( [\mathcal{H},S_{\bot ,\Vert
}]^{2}\right) /Tr\left( S_{\bot ,\Vert }{}^{2}\right) $ and $M_{4}^{\Vert
,\bot }=Tr\left( [\mathcal{H},[\mathcal{H},S_{\bot ,\Vert }]]^{2}\right)
/Tr\left( S_{\bot ,\Vert }{}^{2}\right) $ are the second and fourth moments
respectively, and $S_{\bot ,\Vert }$ stands for the spin component
perpendicular or parallel to the applied field respectively. For the
Hamiltonian of Eq.~\ref{Hamiltonian}, on a kagom\'{e} lattice, and for each
of the field orientation, we obtain the second and fourth moments as given
in the supplementary material. When taking $F=0$ the second moments are
given by 
\begin{align}
M_{2}^{\perp }=& 4E^{2}  \label{SecondMoment} \\
M_{2}^{||}=& E^{2}+D^{2},
\end{align}%
and the fourth moments up to second order in anisotropies are given by 
\begin{align}
M_{4}^{\perp }=& 18J^{2}E^{2}  \label{ForthMoment} \\
M_{4}^{||}=& \frac{9}{2}J^{2}E^{2}+3J^{2}D^{2}
\end{align}%
%
%
%
%
%Details of this calculation are also given in the supplementary material. 
We numerically solve Eqs.~\ref{delta} and \ref{JplusD}. All the possible
solutions of these equations are given in Table~\ref{tab:solu}. 
\begin{table}[h]
\begin{tabular}{l|c|c|c}
\hline
Sol.~\# & J (K) & D (K) & E (K) \\ \hline
1 & -2.3822 & 0.3421 & $\pm$ 0.14217 \\ 
2 & -1.7470 & -0.2930 & $\pm$ 0.12175 \\ \hline
\end{tabular}%
\caption{The Hamiltonian parameters derived from the solution of Eq.~\protect
\ref{delta} and~\protect\ref{JplusD}.}
\label{tab:solu}
\end{table}
In both solutions $J+D<0$ and $J\pm E<0$ and, as stated above, the
interactions between spins are ferromagnetic in all directions.

To check our conclusion, and to verify the assumption made so far, we use
the calculated Hamiltonian parameters to generate the expected field
dependent magnetization in the perpendicular direction $\left\langle M_{\bot
}\right\rangle (H,J,E,g_{\bot })$ and compare it to the data in Fig.~\ref%
{fig:magnet}. For this purpose we calculate the two component magnetization
of each of the three kagom\'{e} sublattices $\mathbf{M}^{d}$ for $\mathbf{H}%
\perp \mathbf{\hat{c}}$, by solving the six coupled explicit Eqs.~\ref%
{Magnetization}. The $\mathbf{H}_{\mathrm{eff}}^{d}$ expressions are given
in the supplementary material. We then average the magnetization on the
three sublattices and project the result onto the $\mathbf{\hat{H}}$
direction to generate $\left\langle M_{\bot }\right\rangle $ as measured
experimentally. For all applied field values, the result is independent of
the applied field direction in the plane. The g-factor determined from the
high field data is $g_{\bot }=2.55(4)$. \ In Fig.~\ref{fig:magnet} we show
the calculated $\left\langle M_{\bot }\right\rangle $ for the two possible
sets of parameters from Table~\ref{tab:solu}. Only solution 1 in the table
agrees with the data. The agreement with the data is nearly prefect.
Therefore, one can fit all our data without the need to introduce $F$.

We attempt to confirm these results by temperature dependent susceptibility $%
\chi \equiv M/H$ measurements for $H\rightarrow 0$. The inverse
susceptibility $\chi ^{-1}$ as a function of temperature, for both field
orientations with an applied field of $0.1$~kG, is depicted in Fig.~\ref%
{fig:suscep}. $\chi ^{-1}$ is clearly different between the two directions.
We performed a high temperature fit to the inverse Curie-Weiss (CW) law, $%
\chi (T)^{-1}=(T-\Theta _{\text{CW}})/C$, where $C$ is the Curie constant,
and $\Theta _{\text{CW}}$ is the CW temperature. For the two experiments the
fit was applied in two temperature ranges: low-$T$ [$5~$K, $30~$K] and high-$%
T$ [$50~$K, $100~$K]. Above 100~K $\chi (T)^{-1}$ is no longer linear with $T
$ for both directions. The CW temperature for $\mathbf{H}\parallel \mathbf{%
\hat{c}}$ from the high-$T$ range is $-1.0(8)$~K, and from the low-$T$ range
is $\Theta _{\text{CW}}^{||}=4.03$~K. The CW temperature for $\mathbf{H}%
\perp \hat{c}$ from the high-$T$ is $\Theta _{\text{CW}}^{\bot }=-49(2)$~K.
For low-$T$, $\mathbf{H}\perp \hat{c}$, the data is not linear and could not
be fitted reliably. The inset of Fig.~\ref{fig:suscep} displays the
difference between the fitted curve and the experimental data. It is clear
that above $\approx 15$~K, this difference for $(\chi (T)^{\perp })^{-1}$
deviates greatly from $0$, whereas the difference for $(\chi (T)^{\parallel
})^{-1}$ is close to $0$. This type of analysis provides a reliable $\Theta
_{\text{CW}}$ only when the temperature range used in the fit is much larger
than the CW temperature obtained by the fit. For $\mathbf{H}\perp \mathbf{%
\hat{c}}$ this condition is not obeyed in the high temperature range.
Therefore $\Theta _{\text{CW}}^{\bot }$ is ambiguous. For $\mathbf{H}%
\parallel \hat{c}$ both temperature ranges are valid, but give conflicting
values of $\Theta _{\text{CW}}^{||}$.

The situation is even more confusing when analyzing the Curie constant for
the different directions and temperature ranges. We found that in both
temperature ranges, the Curie constant is substantially different between
the different directions, and much smaller than expected from localized spin 
$1/2$ on each Cu site. We therefore abandon susceptibility measurements as a
mean of characterizing the Hamiltonian.

We now turn to discuss the longitudinal fields (LF) $\mu $SR results. The
data were collected at the M15 surface muon beamline at TRIUMF using a
dilution refrigerator spectrometer. The spectra were gathered at $T=45$~mK
and $T=240$~mK. In the LF-mode the external field is applied along the
initial muon spin direction. When the internal fields fluctuate in space and
time, the muon spin polarization is expected to complete less than one full
oscillation, and then to relax. The frequency of oscillation increases and
the relaxation rate decreases as the field increases. This behavior is
described by the dynamical LF Kubo-Toyabe (DLFKT) function $G(\Delta ,\nu
,t,H_{\text{LF}})$, where $\nu $ is the field fluctuation rate, $\Delta $ is
the static width of the local field distribution, and $H_{\text{LF}}$ is the
applied field~\cite{Hayano:1979wt}.

Figure \ref{fig:musr}(a) shows the spectra obtained with different fields at 
$T=45$~mK. The data exhibits a typical DLFKT behavior in every respect. We
fit the function 
\begin{equation}
A(t)=A_{0}G(\Delta ,\nu ,t,H_{\text{LF}})+B_{g}  \label{eq:kt}
\end{equation}%
to the data where $A(t)$ is the muon asymmetry, and $B_{g}$ is a
non-relaxing background due to muons stopping in the sample holder. All the
fit parameters are shared for all the data sets at a given temperature. The
instantaneous internal field distribution is assumed to be Gaussian. The fit
is demonstrated by the solid lines in Fig \ref{fig:musr}(a). We obtain that $%
\nu =0.43(2)\mu $s$^{-1}$ and $\Delta =19.3450(4)$~MHz. The value for $%
\Delta $ is consistent with previous measurements~\cite{Marcipar:2009kh}
indicating the same field distribution from the millikelvin to few Kelvin
range. In contrast, $\nu $ decreases by a factor of $\approx 8$ relative to
data obtained before at a temperature 20 times larger ($0.9$~K) \cite%
{Marcipar:2009kh}. We add the new $\nu $ values to the previous results in
Fig.~\ref{fig:musr}(b). The full picture clearly shows dramatic slowing of
the spin fluctuations below $\approx 1.8$~K. However the system continues to
fluctuate even at $0.05$~K with no signs of freezing. Between $240$~mK and $%
45$~mK $\nu $ is finite, clearly measurable by $\mu $SR, and temperature
independent.

It should be pointed out that the analysis of the $\mu $SR data was done
assuming that the muon experiences only the external field. However, in a
ferromagnet the internal field is larger than the external field.
Unfortunately, without proper knowledge of the muon stopping site it is
difficult to estimate the internal field. Analysis of the our data with a
field larger than $H_{\text{LF}}$ could only lead to higher values of $\nu $%
. Therefore, the $\nu $ in Fig.~\ref{fig:musr}(b) should be considered as
lower limit on the real values.

Finite fluctuation rate $\nu $ at $T\rightarrow 0$, with different time
scales, was observed in many kagom\'{e} lattices with antiferromagnetic
interactions such as SCGO \cite{Uemura:1994tz}, Volborthite \cite%
{FukayaPRL91,Bert:2004ij}, Herbertsmithite \cite{Mendels:2007hz,Ofer:2006tj}%
, Nd$_{3}$Ga$_{5}$SiO$_{14}$\cite{Zorko:2008cc}, Langasite \cite{Zorko:
2010jr}, Kapellasite \cite{Fak:2012iu}, and vanadium-oxyfluoride \cite%
{Clark:2013ih}. All these compounds are considered to be SL. However the
Hamiltonian in Eq.~\ref{Hamiltonian}, with the parameters in Table~\ref%
{tab:solu}, gives a ground state that is fundamentally different from these
spin liquids. Since $D>\left\vert E\right\vert $ the spin lie in the xy
plane. This allows us to define one angle per spin as shown in the inset of
Fig. 4(b). For positive or negative $E$ the spins would like to lie parallel
or perpendicular to a bond, respectively. However, the ground state energy
minimum is reached when two spins make the angles $\alpha =-\beta =-\sqrt{3}%
E/(6J-E)$ with a bond, and the third spin has $\gamma =0$ and is $60$
degrees away from a bond (see supplementary material). This is a slightly
frustrated spin arrangement. A new energy minimum for all the spins on the
lattice is found every $60$ degrees, but there is no local continuous
degeneracy. The energy minimum is shallow and it takes $\sim 10$~mK per unit
cell to overcome the potential barrier and move the entire spin system
collectively between local energy minima. This means that Cu(1,3-bdc) should
order magnetically and it is not a spin liquid.

In summary, the Cu(1,3-bdc), with Cu$^{+2}$ spin-$1/2$ situated on kagom\'{e}
lattice, exhibits anisotropic but ferromagnetic interactions in all
direction. $\mu $SR indicates persistent spin dynamics down to $45$~mK as
expected from a spin liquid. The same behavior was observed in many kagom%
\'{e} lattices with AFM interactions. This is very surprising given that a
kagom\'{e} lattice with ferromagnetic interactions has very small degree of
frustration, lacks continuous local degeneracy, and is not a spin liquid.
Therefore, $\mu $SR can falsely identify a spin liquid.

This work was supported by the Israel USA binational science foundation. We
would like to thank Young S. Lee and Joel S. Helton for providing us with
the samples. The authors wishes to thank the TRIUMF staff for help with the $%
\mu $SR experiments. Helpful discussions with Sarah Dunsiger are greatly
acknowledged.

\section{Supplementary Material}

\subsection{Moments}

The evaluation of the linewidths in ESR at high temperatures involves
calculating the second and fourth order moments to be used in Eq.~6 in the
main text. Thus we need to evaluate the first and second order commutators
of the total spin component in a given direction and the Hamiltonian. It is
straightforward to see that for a spin-$1/2$ Hamiltonian the first order
commutator gives rise to two-spin terms. The second order commutator gives
rise to single spin or three spin terms depending on whether a bond term,
from the Hamiltonian, and a two spin term from the first order commutator,
share two sites or one.\newline

The evaluation of the traces involves careful bookeeping of all the
different terms possible. We accomplished that by writing a Mathematica
program which evaluates all the terms that can arise and computes the trace.
The program evaluates the moments for a general Hamiltonian of spin-$1/2$
sites. It is assumed that the lattice can be grouped into clusters of sites,
which can be seen as sublattices corresponding to a particular Bravais
lattice site. The Hamiltonian is then specified as the sum of interactions
within a cluster and between clusters. Thus we assume that the Hamiltonian
has translation invariance. The input to the program specifies all the site
indices and coupling coefficients ($9$ in number, for $S_{i\alpha }S_{j\beta
}$) for all the different bonds involving spins from the Bravais lattice
site at the origin. The rest of the bonds on the lattice and their
contributions to the trace can be evaluated given the translation
invariance. Thus the evaluation is quite general and can be extended to
several other systems with more general Hamiltonians such as those
containing all the components of the DM interaction and also longer range
exchange interactions.

The moments in this paper have been evaluated for the anisotropic kagom\'{e}
Hamiltonian with three exchange coupling constants and a
Dzyaloshinski-Moriya (DM) term given in Eq.~1 of the main text. They are
given by: 
\begin{align}
M_{2}^{\perp }=& 4E^{2}  \label{SecondMomentFull} \\
M_{2}^{||}=& F^{2}+E^{2}+D^{2} \\
M_{4}^{\perp }=& 18J^{2}E^{2}+28E^{4}+10F^{2}E^{2}+8\sqrt{3}FE^{2}J  \notag
\\
& +4\sqrt{3}FE^{2}D+20JDE^{2}+8E^{2}D^{2}  \label{ForthMomentFull} \\
M_{4}^{||}=& \frac{11}{2}F^{4}-JDE^{2}+\frac{9}{2}J^{2}E^{2}+3J^{2}D^{2} \\
& +2JD^{3}-\frac{5\sqrt{3}}{2}FE^{2}D+\frac{13}{4}E^{4}+\frac{5}{2}D^{4} 
\notag \\
& +\frac{41}{4}F^{2}E^{2}+2F^{2}J^{2}+3F^{2}D^{2}+\frac{25}{4}E^{2}D^{2} 
\notag
\end{align}

\subsection{Effective Fields}

The kagom\'{e} lattice is constructed from three sublattices. They are
presented in Fig.~\ref{fig:musr}(b) of the paper. The effective fields in
the three different sublattices are:%
\begin{eqnarray}
\mathbf{H}_{\text{eff}}^{a} &=&\mathbf{H}-\frac{2J}{g^{2}\mu _{B}^{2}}(%
\mathbf{M}^{b}+\mathbf{M}^{c})-\frac{E}{g^{2}\mu _{B}^{2}}\Big(2M_{x}^{b} \\
&&-M_{x}^{c}+\sqrt{3}M_{y}^{c},-2M_{y}^{b}+M_{y}^{c}+\sqrt{3}M_{x}^{c}\Big) 
\notag \\
\mathbf{H}_{\text{eff}}^{b} &=&\mathbf{H}-\frac{2J}{g^{2}\mu _{B}^{2}}(%
\mathbf{M}^{a}+\mathbf{M}^{c})-\frac{E}{g^{2}\mu _{B}^{2}}\Big(2M_{x}^{a} \\
&&-M_{x}^{c}-\sqrt{3}M_{y}^{c},-2M_{y}^{a}+M_{y}^{c}-\sqrt{3}M_{x}^{c}\Big) 
\notag \\
\mathbf{H}_{\text{eff}}^{c} &=&\mathbf{H}-\frac{2J}{g^{2}\mu _{B}^{2}}(%
\mathbf{M}^{b}+\mathbf{M}^{a})-\frac{E}{g^{2}\mu _{B}^{2}}\Big(-M_{x}^{a} \\
&&-M_{x}^{b}+\sqrt{3}M_{y}^{a}-\sqrt{3}M_{y}^{b},M_{y}^{a}+M_{y}^{b}+\sqrt{3}%
M_{x}^{a}-\sqrt{3}M_{x}^{b}\Big)  \notag
\end{eqnarray}

\subsection{Ground state}

We determine the ground state in the mean field approximation by writing%
\begin{eqnarray}
\mathbf{M}^{a} &=&M(\cos \alpha ,\sin \alpha ) \\
\mathbf{M}^{b} &=&M(\cos \beta ,\sin \beta ) \\
\mathbf{M}^{c} &=&M(\cos \gamma ,\sin \gamma ).
\end{eqnarray}%
The Hamiltonian per unit cell $H=-\frac{1}{2}\left( \mathbf{M}^{a}\cdot 
\mathbf{H}_{eff}^{a}+\mathbf{M}^{b}\cdot \mathbf{H}_{eff}^{b}+\mathbf{M}%
^{c}\cdot \mathbf{H}_{eff}^{c}\right) $ in zero external field it is given by%
\begin{eqnarray}
H &=&\frac{M^{2}}{2(g\mu _{B})^{2}}\{J\left[ 4\cos (\alpha -\beta )+4\cos
(\beta -\gamma )+4\cos (\gamma -\alpha )\right] +  \notag \\
&&E[4\cos (\alpha +\beta )+2\sqrt{3}\sin (\alpha +\gamma )-2\sqrt{3}\sin
(\beta +\gamma ) \\
&&-2\cos (\alpha +\gamma )-2\cos (\beta +\gamma )]\}  \notag
\end{eqnarray}%
This Hamiltonian is invariant under rotations by $120$ degrees and cyclic
permutations of the angles. A numerical search for the minimum shows that it
occurs at $\alpha =-\beta $, and $\gamma =0$. Given these relations, and
that for each sublattice $\mathbf{H}_{\text{eff}}^{d}$ is parallel to $%
\mathbf{M}^{d}$, we find that $\alpha =-\beta =-\sqrt{3}E/(6J-E)$.


\begin{thebibliography}{99}
\bibitem{Uemura:1994tz} Y. J. Uemura, A. Keren, K. Kojima, L. P. Le, G. M.
Luke, W. D. Wu, Y. Ajiro, T. Asano, Y. Kuriyama, M. Mekata, H. Kikuchi, and
K. Kakurai, Phys. Rev. Lett. \textbf{73} 3306 (1994).

\bibitem{FukayaPRL91} A. Fukaya, Y. Fudamoto, I. M. Gat, T. Ito, M. I.
Larkin, A. T. Savici, Y. J. Uemura, P. P. Kyriakou, G. M. Luke, M. T.
Rovers, K. M. Kojima, A. Keren, M. Hanawa, and Z. Hiroi, Phys. Rev. Lett. 
\textbf{91}, 207603 (2003)

\bibitem{Bert:2004ij} F. Bert, D. Bono, P. Mendels, J-C. Trombe, P. Millet,
A. Amato, C. Baines, and A. Hillier, J. Phys.: Condens. Matter \textbf{16}
S829 (2004).

\bibitem{Mendels:2007hz} P. Mendels, F. Bert, M. A. de Vries, A. Olariu, A.
Harrison, F. Duc, J. C. Trombe, J. S. Lord, A. Amato, and C. Baines, Phys.
Rev. Lett. \textbf{98}, 077204 (2007).

\bibitem{Ofer:2006tj} Oren Ofer and Amit Keren, E. A. Nytko, M. P. Shores,
B. M. Bartlett, D. G. Nocera, C. Bains and A. Amato, unpublished
arXiv:cont-mat/0610540v2 (2006).

\bibitem{Zorko:2008cc} A. Zorko, F. Bert, P. Mendels, P. Bordet, P. Lejay,
and J. Robert, Phys. Rev. Lett. \textbf{100}, 147201 (2008).

\bibitem{Zorko: 2010jr} A. Zorko, F. Bert, P. Mendels, P. Bordet, P. Lejay,
and J. Robert, Phys. Rev. Lett. \textbf{100}, 057202 (2008).

\bibitem{Fak:2012iu} B. F\aa k, E. Kermarrec, L. Messio, B. Bernu, C.
Lhuillier, F. Bert, P. Mendels,5, B. Koteswararao, F. Bouquet, J. Ollivier,
A. D. Hillier, A. Amato, R. H. Colman, and A. S. Wills, Phys. Rev. Lett. 
\textbf{109}, 037208 (2012).

\bibitem{Clark:2013ih} L. Clark, J. C. Orain, F. Bert, M. A. de Vries, F. H.
Adidoudi, R. E. Morris P. Lightfoot, J. S. Lord, M. T. F. Telling, P.
Bonville, J. P. Attfield, P. Mendels, and A. Harrison, Phys. Rev. Lett. 
\textbf{110}, 207208 (2013).

\bibitem{DalmasdeReotier:2006kj} P. Dalmas de Reotier, A. Yaouanc, L.
Keller, A. Cervellino, B. Roessli, C. Bains, A. Forget, C. Vaju, P. C. M.
Gubbens, A. Amato, and P. J. C. King, Phys. Rev. Lett. \textbf{96} 127202
(2006).; A. Yaouanc, P. Dalmas de Reotier, P. Bonville, J. A. Hodges, V.
Glazkov, L. Keller, V. Silkolenko, M. Bartkowiak, A. Amato, C. Bains, P. J.
C. King, P. C. M. Gubbens, and A. Forget, Phys. Rev. Lett. \textbf{110},
127207 (2013).; J. S. Gardner, S. R. Dunsiger, B. D. Gaulin, M. J. P.
Gingras, J. E. Greedan, R. F. Kiefl, M. D. Lumsden, W. A. MacFarlane, N. P.
Raju, J. E. Sonier, I. Swainson, and Z. Tun, Phys. Rev. Lett. \textbf{82}
1012 (1999).; A. Keren, J. Gardner, G. Ehlers, A. Fukaya, E. Segal, and Y.
Uemura, Phys. Rev. Lett. \textbf{92}, 107204 (2004).;

\bibitem{SalmanPRB} Zaher Salman, Amit Keren, Philippe Mendels, Valerie
Marvaud, Ariane Scuiller, Michel Verdaguer, James S. Lord, and Chris Baines,
Phys. Rev. B 65, 132403 (2002).

\bibitem{pratt} F. L. Pratt, P. J. Baker, S. J. Blundell, T. Lancaster, S.
Ohira-Kawamura, C. Baines, Y. Shimizu, K. Kanoda, I. Watanabe and G. Saito,
Nature \textbf{471}, 612 (2011).

\bibitem{Nytko:2008ge} Emily A. Nytko, Joel S. Helton, Peter Muller and D.
G. Nocera, J. Am. Chem. Soc. \textbf{130}, 2922 (2008).

\bibitem{Marcipar:2009kh} Lital Marcipar, Oren Ofer, Amit Keren, Emily A.
Nytko, Daniel G. Nocera, Young S. Lee, Joel S. Helton, Chris Bains, Phys.
Rev. B. {80}, 132402 (2009).

\bibitem{ElhahalPRB02} M. Elhahal, B. Canals, and C. Lacroix, Phys. Rev. B 
\textbf{66}, 014422 (2002).

\bibitem{ZorkoPRL08} A. Zorko, S. Nellutla, J. van Tol, L. C. Brunel, F.
Bert, F. Duc, J.-C. Trombe, M. A. de Vries, A. Harrison, and P. Mendels,
Phys. Rev. Lett. \textbf{101}, 026405 (2008). 
%\bibitem{} A. Harrison,5 and P. Mendels1

\bibitem{Ofer:2009hy} Oren Ofer and Amit Keren, Phys. Rev. B \textbf{79},
134424 (2009).

%\bibitem{Ravi} Ravi, in preparation.

\bibitem{Hayano:1979wt} R. S. Hayano, Y. J. Uemura, J. Imazato, N. Nishida,
T. Yamazaki, and R. Kubo, Phys. Rev. B \textbf{20}, 850 (1979).
\end{thebibliography}
\end{document}